\documentclass[twocolumn,pra,showpacs,preprintnumbers,amsfonts,amsmath,amssymb,superscriptaddress]{revtex4}

\usepackage{graphicx}
\usepackage{dcolumn}
\usepackage{bm}

\usepackage[bookmarks=false,pdfstartview={FitH}]{hyperref}

\newcommand{\su}{\mathfrak{su}}
\newcommand{\SU}{\mathrm{SU}}

\newcommand{\U}{\mathrm{U}}

\newcommand{\R}{\mathbb{R}}
\newcommand{\Z}{\mathbb{Z}}

\newcommand{\G}{\mathrm{G}}
\newcommand{\GA}{\mathrm{A}}

\newcommand{\GK}{\mathrm{K}}

\newcommand{\g}{\mathfrak{g}}

\newcommand{\gk}{\mathfrak{k}}
\newcommand{\gp}{\mathfrak{p}}
\newcommand{\ga}{\mathfrak{a}}

\newcommand{\idt}{\mathrm{id}_{2}}
\newcommand{\idf}{\mathrm{id}_{4}}
\newcommand{\idd}{\mathrm{id}_{d}}

\newcommand{\We}{\mathcal{W}}

\providecommand{\abs}[1]{\lvert#1\rvert}

\begin{document}

\title{Time-optimal synthesis of unitary transformations\\
in coupled fast and slow qubit system}

\author{Robert Zeier}%
\email{zeier@eecs.harvard.edu}
\affiliation{%
Harvard School of Engineering and Applied Sciences,
33 Oxford Street,
Cambridge, Massachusetts 02138, USA
}%
\author{Haidong Yuan}%
\email{haidong@mit.edu}
\affiliation{%
Department of Mechanical Engineering,
Massachusetts Institute of Technology,
77 Massachusetts Avenue,
Cambridge, Massachusetts 02139, USA
}%
\author{Navin Khaneja}%
\email{navin@hrl.harvard.edu}
\affiliation{%
Harvard School of Engineering and Applied Sciences,
33 Oxford Street,
Cambridge, Massachusetts 02138, USA
}%

\date{September 27, 2007}

\begin{abstract}
In this paper, we study time-optimal control problems related to
system of two coupled qubits where the time scales involved in performing 
unitary transformations on each qubit are significantly different.
In particular, we address the case where unitary transformations produced by evolutions
of the coupling take much longer time as compared to the time required to produce
unitary transformations on the first qubit but much shorter time as compared to the time
to produce unitary transformations on the second qubit.
We present a 
canonical decomposition
of $\SU(4)$ in terms of the subgroup $\SU(2)\times\SU(2)\times\U(1)$, which is natural in 
understanding the time-optimal control problem
of such a coupled qubit system with significantly different time scales.
A typical setting involves dynamics of a coupled electron-nuclear spin
system in pulsed electron paramagnetic resonance experiments at high fields. 
Using the proposed canonical decomposition, we give time-optimal control algorithms to synthesize 
various unitary transformations of interest in coherent spectroscopy and quantum information
processing.
\end{abstract}

\pacs{03.67.Lx}

\maketitle
\section{Introduction}
The synthesis of unitary transformations using time-efficient control algorithms 
is a well studied problem
in quantum information processing and coherent spectroscopy. Time-efficient control algorithms
can reduce decoherence effects in experimental realizations, and the study
of such control algorithms is related to the complexity
of quantum algorithms (see, e.g., \cite{NDGD:2006,Nie:2006,NDGD:2006b}).
Significant literature in this subject treat the case
where unitary transformations on single qubits take
negligible time compared
to transformations interacting between different qubits. 
This particular assumption is very realistic for nuclear spins in nuclear magnetic resonance (NMR) spectroscopy.
Under this assumption,
Ref.~\cite{KBG:2001}  (see also 
\cite{KG:2001b,RKG:2002,VHC:2002,HVC:2002,CHN:2003,ZGB:2004,KKG:2005,YK:2005,YK:2005b,Swo:2006,YK:2006,Yua:2006,DHHKL:2006,Zei:2006}) presents time-optimal control algorithms to synthesize
arbitrary unitary transformations on a system of two qubits.
Further progress in the case of multiple 
qubits is reported in \cite{KG:2001b,KGB:2002,RKG:2003,ZGB:2004,KRKSG:2005,SSK:2005,Yua:2006,Zei:2006,YK:2006b,KHSY:2007,YGK:2007}.

In this work, we consider a coupled qubit system where local unitary transformations on the 
first qubit take significantly less time than local transformations on the second one. In addition, we
assume that the coupling evolution  is much slower than transformations on the first qubit but
much faster than transformations on the second one. We present a canonical decomposition of $\SU(4)$
in terms of the subgroup $\SU(2)\times\SU(2)\times\U(1)$ reflecting the significantly different time scales immanent in the system.
Employing this canonical decomposition, we derive  
time-optimal control algorithms
to synthesize various unitary transformations.
Our methods are applicable to coupled electron-nuclear spin systems occurring
in pulsed electron paramagnetic resonance (EPR) experiments at high fields, where the 
Rabi frequency of the electron is much larger than
the hyperfine coupling which is further
much larger than the Rabi frequency of the nucleus.
In the context of quantum computing similar
electron-nuclear spin systems appear in the Refs.~\cite{MMS:2003,JGP:2004a,JGP:2004,MSW:2004,Men:2005,HMM:2006,Rah:2006,MM:2006,GDP:2006,MHS:2006,HMM:2006b,JW:2006,WJ:2006,CGDT:2006,Hei:2006,AF:2007,GDCJ:2007}. 

The main results of this paper are as follows.  Let $S_{\mu}$ and $I_{\nu}$ represent 
spin operators for the fast (electron spin) and slow (nuclear spin) qubit, respectively.
Any unitary transformation $G \in \SU(4)$ on the coupled spin system 
can be decomposed as 
\begin{equation}
\label{eq:main00}
G = K_1 \exp(t_1 S^{\beta}I_x + t_2 S^{\alpha}I_x )K_2,
\end{equation}
where $S^{\alpha}I_x$ and $S^{\beta}I_x$ correspond to
$x$-rotations of the slow qubit, conditioned, respectively, on the up or 
down state of the fast qubit. 
The elements $K_1$ and $K_2$ are rotations synthesized by rapid
manipulations of the fast qubit in conjunction with the evolution of 
the natural Hamiltonian. The elements $K_1$ and $K_2$ 
belong to the subgroup $\SU(2) \times \SU(2) \times \U(1)$, and in 
appropriately chosen basis correspond to block-diagonal special unitary 
matrices with  $2{\times}2$-dimensional blocks of unitary matrices. 

The minimum time to produce any unitary transformation $G$ is the smallest value of
$(\abs{t_1} + \abs{t_2})/{\omega_r^I}$, where 
$\omega_r^I$ is the maximum achievable Rabi frequency 
of the nucleus and $(t_1, t_2)^{T}$ is a pair 
satisfying Eq.~\eqref{eq:main00}. Synthesizing $K_1$ and $K_2$ takes 
negligible time on the time scale governed by $\omega_r^I$.

The paper is organized as follows.
In Sec.~\ref{physicalmodel}, we recall the physical details of our model
system exemplified by a coupled electron-nuclear spin system.
The Lie-algebraic structure of our model is described in Sec.~\ref{structure}, which 
is used to derive control algorithms (pulse sequences) for synthesizing arbitrary 
unitary transformations in our coupled spin system.
In Sec.~\ref{examples}, we present examples.
We prove the time-optimality of our
control algorithms in Sec.~\ref{derivation}, and some details of the proof
are given in  
Appendix~\ref{proofs}.

Our work draws some results from the theory of Lie groups, which are explained as 
needed. We refer to \cite{Hel:2001,Kna:2002} for general reference.
To make the paper broadly accessible, we work with explicit 
matrix representations of Lie groups and Lie algebras.

\section{Physical model\label{physicalmodel}}
As our model system, we consider two coupled qubits.
We introduce the operators  $S_{\mu}$ and $I_{\nu}$ 
which correspond to operators on the first and second qubit, respectively.
In particular, these operators are defined by $S_{\mu}=(
\sigma_{\mu}\otimes\mathrm{id}_2)/2$
and $I_{\nu}= (\mathrm{id}_2\otimes\sigma_{\nu})/2$  (see \cite{EBW:1997}), where
$\sigma_{x}:=
\left(\begin{smallmatrix}
0 & 1 \\
1 & 0
\end{smallmatrix}
\right)$,
$
\sigma_{y}:=
\left(
\begin{smallmatrix}
0 & -i \\
i & 0
\end{smallmatrix}
\right)
$, and
$
\sigma_{z}:=
\left(
\begin{smallmatrix}
1 & 0 \\
0 & -1
\end{smallmatrix}
\right)
$ are the Pauli matrices
and
$
\idt:=
\left(
\begin{smallmatrix}
1 & 0 \\
0 & 1
\end{smallmatrix}
\right)
$
is the $2{\times}2$-dimensional identity matrix. 
In the remaining text, let $\mu, \nu \in \{x,y,z\}$ and $\gamma \in \{x, y \}$.

In an experimental setting using an electron-nuclear spin system,
the first qubit is represented by the electron spin (of spin $1/2$).
Similarly, the second qubit is represented by the nuclear spin (of spin $1/2$).
We assume that in the presence of a static magnetic field pointing in the $z$-direction, 
the free evolution is governed in the lab frame by
a Hamiltonian of the form
\begin{equation}
\label{eq:lab}
H_{0}^{\text{lab}} = \omega_S S_z + \omega_I I_z + J (2 S_z I_z),
\end{equation}
where $\omega_S$ and $\omega_I$ represents the natural precession frequency of, respectively, the first qubit
and second qubit and $J$  is the coupling strength. We assume that
\begin{equation}
\label{eq:scales}
\omega_{S} \gg \omega_{I} \gg J.
\end{equation} This assumption is 
motivated by coupled 
electron-nuclear spin system
occurring in EPR experiments at high fields (see, e.g., Sect.~3.5 of \cite{SJ:2001}).
The time scales in Eq.~\eqref{eq:scales} insure that the hyperfine coupling Hamiltonian 
between the spins averages to the Ising Hamiltonian $2S_z I_z$, as in Eq.~\eqref{eq:lab}. This is 
the so-called high field limit.

The first and second qubit are controlled by transverse oscillating fields, which 
result in the corresponding control Hamiltonian given by
$H_{S}^{\text{lab}}+H_{I}^{\text{lab}}$,
where
\begin{equation*}
H_{S}^{\text{lab}}=2 \omega_{r}^{S}(t) \cos[\omega_{c}^{S} t + \phi_{S}(t)] S_{x}
\end{equation*}
is the control Hamiltonian of the first qubit and
\begin{equation}\label{controlI}
H_{I}^{\text{lab}}=2 \omega_{r}^{I}(t) \cos[\omega_{c}^{I} t + \phi_{I}(t)] I_{x}
\end{equation}
is the control Hamiltonian of the second qubit.
The amplitude, frequency, and phase 
of the control function w.r.t.\ the first qubit are represented by $\omega_{r}^{S}(t)$,
$\omega_{c}^{S}$, and
$\phi_{S}=\phi_{S}(t)$ respectively. Similarly, $\omega_{r}^I(t)$,  $\omega_{c}^{I}$,  and $\phi_{I}=\phi_{I}(t)$ 
represents the amplitude, frequency, and phase of the control function w.r.t.\ the second qubit.
We use $\omega_r^I$ and $\omega_r^S$ to denote the maximal possible values of 
$\omega_r^I(t)$ and $\omega_r^S(t)$. 
In our model system, we assume that 
\begin{equation} 
\label{timescales}
\omega_{r}^I \ll J \ll \omega_{r}^S.
\end{equation}
Therefore, we refer to the first qubit as the fast qubit and the second qubit as the slow qubit. 

We choose $\omega_{c}^{S}=\omega_S$ and $\omega_{c}^{I}=\omega_I - J$. In a double rotating 
frame, rotating with the first and second qubit at frequency 
$\omega_{c}^{S}$ and $\omega_{c}^{I}$, the transformations
$U_{\text{lab}}(t)$ and $U_{\text{rot}}(t)$ describe, respectively, a
unitary transformation in the lab frame and the double rotating frame.
We have
$$
U_{\text{lab}}(t)=\exp(-i t\omega_{c}^{S} S_{z}) \exp(-i t \omega_{c}^{I} I_{z}) U_{\text{rot}}(t).
$$
Using the rotating wave approximation, the Hamiltonians $H_{0}^{\text{lab}}$, $H_{S}^{\text{lab}}$, and 
$H_{I}^{\text{lab}}$ transform, respectively, to 
\begin{align}\label{eq:H0}
H_{0} =\, & J I_{z} + J (2S_{z}I_{z}), \\
\nonumber H_{S} =\, & \omega_{r}^S(t) [S_x \cos \phi_{S}(t)+  S_y \sin \phi_{S}(t)], \\ \intertext{and}
\nonumber H_{I} =\, & \omega_{r}^I(t) [I_x \cos \phi_{I}(t)+  I_y \sin \phi_{I}(t)].
\end{align}
In absence of any irradiation on qubits, the system evolves under the free Hamiltonian $-iH_0$. 
From the time scales in Eq.~\eqref{timescales}, we can synthesize 
any unitary transformation of the 
form $\exp(-it S_{\mu})$ in arbitrarily small time as compared to 
the evolution under $H_0$ or $H_{0}+H_I$.

Let us define the operators, 
$$
S^{\beta} = (\idf/2  + S_z) = \begin{pmatrix}
\idt & 0_{2} \\
0_{2} & 0_{2}
\end{pmatrix}
$$ 
and 
$$
S^{\alpha} = (\idf/2  - S_z)= \begin{pmatrix}
0_{2} & 0_{2} \\
0_{2} & \idt
\end{pmatrix},
$$
where $\idd$ is the $d{\times}d$-dimensional identity matrix and
$0_2$ is the  $2{\times}2$-dimensional
zero matrix.
Note that $H_0 = 2J S^{\beta}I_z$, and the system is described by the
Hamiltonian $$ H_{0} +H_{I} = 2 J S^{\beta} I_z + w_r^I(t)(S^{\alpha} + S^{\beta})(I_x \cos\phi_{I} + I_y \sin\phi_{I}). $$ 
Since $J \gg w_r^I(t) $, and $S^{\beta} I_{\gamma}$, does not commute with $S^\beta I_z$, 
the above Hamiltonian gets in the first order approximation truncated to
\begin{equation}
\label{eq:alpha}
H^\alpha(\phi_{I}) = 2J S^{\beta} I_z + w_r^I(t)S^{\alpha} (I_x \cos \phi_{I} + I_y \sin \phi_{I} ).
\end{equation}
Similarly, we can prepare an Hamiltonian
\begin{equation}
\label{eq:beta} 
 H^\beta(\phi_{I}) = 2J S^{\alpha} I_z + w_r^I(t)S^{\beta} (I_x \cos \phi_{I} + I_y \sin \phi_{I} )
\end{equation} 
by using
$ H^\beta(\phi_{I}) = \exp(i\pi S_x) H^\alpha(\phi_{I}) \exp(-i \pi S_x) $. 

\begin{figure}
\includegraphics{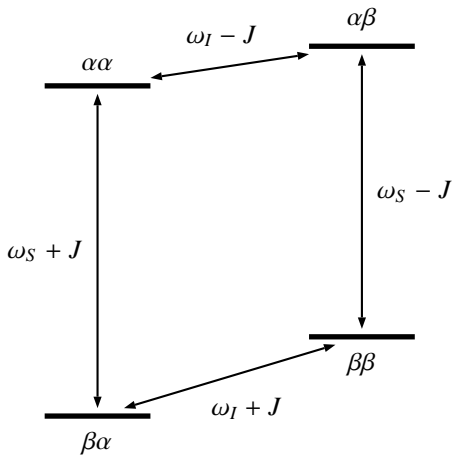}
\caption{\label{fig:energy} The eigenstates of the Hamiltonian $H_{0}^{\text{lab}}$
are shown, where the transitions $\alpha\alpha \leftrightarrow \alpha\beta$ and 
$\beta\alpha \leftrightarrow \beta\beta$
correspond respectively to the orientation along and
opposite to the static magnetic field.
The first and second index refer to the 
orientation of the electron and nuclear spin, respectively.
Refer to the text for details.}
\end{figure}

The Hamiltonians $H^\alpha(\phi_I)$ and $H^\beta(\phi_I)$, operate on the slow qubit 
and induce transitions $\alpha\alpha \leftrightarrow \alpha\beta$ and $\beta\alpha \leftrightarrow \beta\beta$ 
 of the nuclear spin as shown in Fig.~\ref{fig:energy} (cp.\ Table~6.1.1 of \cite{SJ:2001}).
The $\alpha$ and $\beta$ states of the spins denote their orientation along and
opposite to the static magnetic field, respectively.
For the electron spin, the $\beta$ state has lower energy than the $\alpha$ state
as its gyromagnetic ratio is negative. Similarly, for the nuclear spin, the $\alpha$ state has lower energy than the $\beta$ state
as its gyromagnetic ratio is positive (as for a proton).
We remark that the energy eigenstates $\beta \alpha$, $\beta \beta$, $\alpha \alpha$,
and $\alpha \beta$ correspond, respectively, to the basis states $00$, $01$, $10$, and  $11$.
In Fig.~\ref{fig:energy}, the first and second 
index in eigenstates refers to the orientation of the electron and nuclear spin, respectively.
In absence of any irradiation on the two qubits, the system evolves under the Hamiltonian $-iH_0$.
In this section, we have shown how to synthesize generators of the form 
$-iS_\mu$, $-iH^\alpha(\phi_{I})$, and $-iH_0$.

\section{Lie-algebraic structure of the model system\label{structure}}
All transformations of our model system are contained in the Lie group
$\G=\SU(4)$, which is the set of  $4{\times}4$-dimensional unitary transformations 
of determinant one. The operators
$-iI_{\mu}$, $-iS_{\nu}$, and $-i2I_{\mu}S_{\nu}$, are 
infinitesimal generators of the Lie group $\G$, and they generate the 
15-dimensional Lie algebra $\g=\su(4)$ given by the (real) vector space of $4{\times}4$-dimensional
(traceless) skew Hermitian matrices. We have shown how to synthesize generators of the form
$-iS_\mu$, $-iH^\alpha(\phi_{I})$, and $-iH_0$. These generators are sufficient to 
produce any unitary transformation on the coupled qubit system, as described below.

{\it Lemma~1.} The Lie algebra generated by the elements
$-iS_{\mu}$, $-i H^{\alpha}(\phi_{I})$, and $-iH_{0}$, is equal to $\g=\su(4)$. 

Therefore, a standard result on the controllability of
(Thm.~7.1 of Ref.~\cite{JS:1972}) implies that the system is completely controllable 
and any unitary transformation in $\G=\SU(4)$, can be synthesized by alternate 
evolution under the above Hamiltonians.

{\it Lemma~2.}
The Lie algebra $\gk$, generated by the elements
$-iS_{\mu}$ and $-iH_{0}$ consists of the  
elements $-iS_{\mu}$, $-i2S_{\nu}I_{z}$, and $-iI_{z}$.

The Lie algebra $\gk$ represents a class of generators that take significantly less time to be synthesized, as they 
only involve controlled rotations of the fast qubit and evolution of the free 
Hamiltonian $-iH_{0}$ (no controlled rotations of the slow qubit are involved). We can  
decompose 
\begin{equation} \label{Cartan}
\g=\gk\oplus\gp,
\end{equation} 
where the 
subspace $\gp$ (of $\g$) consists of the elements $-iI_{\gamma}$ and $-i2S_{\mu}I_{\gamma}$. 
The decomposition of Eq.~\eqref{Cartan} is a  
 Cartan decomposition (see, e.g., \cite{Hel:2001}, p.~213) as
\begin{equation}
\label{eq:Cartan}
[\gk,\gk]\subset\gk,\; [\gk,\gp]\subset\gp,\; \text{and}\;  [\gp,\gp]\subset\gk,
\end{equation}
where $[g_{1},g_{2}]=g_{1}g_{2}-g_{2}g_{1}$ is the commutator ($g_{i}\in \g$).

Let $\GK= \exp(\gk)$ denote the subgroup of $\G=\SU(4)$ which is infinitesimally generated by $\gk$.
The elements of $\GK$ can be synthesized only by the 
free evolution and employing controlled transformations on the fast qubit.
Therefore, synthesizing transformations of $\GK$ takes significantly less time
as compared to general unitary transformations not contained in $\GK$. In particular,
controlled  transformations on the slow qubit are necessary to synthesize
general unitary transformations. 
The Lie group $\GK=\exp(\gk)$ is equal to
$\mathrm{S}[\U(2)\times\U(2)]$, which is
sometimes referred as $\SU(2) \times \SU(2)\times\U(1)$. 

Consider a maximal Abelian subalgebra $\ga$ contained in $\gp$. In our case,
$\ga$ is spanned by the operators $-iS^{\beta}I_{x}$ and $-i
S^{\alpha}I_{x}$. Any element $a\in\ga$ can be represented as
$a_{1}(-iS^{\beta}I_{x})+a_{2}(-iS^{\alpha}I_{x})$, where 
$a_1,a_2\in\R$. As a matrix, $a$ takes the form
\begin{equation*}
-\frac{i}{2}
\begin{pmatrix}
0 &  a_{1} & 0 & 0\\
a_{1} & 0 & 0 & 0\\
0 & 0 & 0 & a_{2}\\
0 & 0 & a_{2} & 0
\end{pmatrix}.
\end{equation*}
We obtain the Lie group $\GA=\exp(\ga)$ corresponding to the Abelian algebra $\ga$.
From a Cartan decomposition of a real 
semisimple Lie-algebra as satisfying Eqs.~\eqref{Cartan}-\eqref{eq:Cartan}, we obtain 
 a decomposition of the compact Lie group
$\G=\GK \GA \GK$ (see, e.g., \cite{Hel:2001}, Chap.~V, Thm.~6.7):

{\it Lemma~3.}
Any element $G \in \SU(4)$, can be written as 
\begin{equation}
\label{eq:decompose}
G=K_{1} \exp[t_{1}(-iS^{\beta}I_{x})+t_{2}(-iS^{\alpha}I_{x})] K_{2},
\end{equation} where $t_{1},t_{2}\in \R$ and $K_{1},K_{2}\in \GK$. 

{\it Remark~1.} The computation of $\GK \GA \GK$ decompositions 
was analyzed in Refs~\cite{Bul:2004,BB:2004,BBL:2005,EP:2005,NKS:2006}. 
In this work, we consider the Cartan decomposition, which
corresponds to the type AIII in the classification of possible Cartan decompositions
(see, e.g., pp.~451--452 of Ref.~\cite{Hel:2001}).

 Transforming all elements $G \in \G$ to
$\mathrm{SWAP}\cdot G\cdot \mathrm{SWAP}$, where
$$
\mathrm{SWAP}= \exp(-i \pi\, \mathbf{S} \cdot \mathbf{I}) =
\begin{pmatrix}
1 & 0 & 0 & 0 \\
0 & 0 & 1 & 0 \\
0 & 1 & 0 & 0 \\
0 & 0 & 0 & 1 
\end{pmatrix},
$$
$\mathbf{S}=(S_{x},S_{y},S_{z})^{T}$, and $\mathbf{I}=(I_{x},I_{y},I_{z})^{T}$,
the $\GK \GA \GK$ decomposition is given in explicit matrices
by
\begin{multline*}
\begin{pmatrix}
U_1 & 0_2\\
0_2 & U_2
\end{pmatrix}
\exp\left[
-\frac{i}{2}
\begin{pmatrix}
0 & 0 & a_{1}  & 0\\
0 & 0 & 0 & a_{2}\\
a_{1} & 0 & 0 & 0\\
0 & a_{2} & 0 & 0
\end{pmatrix}
\right]
\begin{pmatrix}
U_3 & 0_2\\
0_2 & U_4
\end{pmatrix}\\
=
\begin{pmatrix}
U_1 & 0_2\\
0_2 & U_2
\end{pmatrix}
\begin{pmatrix}
c_{1} & 0 & -i s_{1}  & 0\\
0 & c_{2} & 0 & -i s_{2}\\
-i s_{1} & 0 & c_{1} & 0\\
0 & -i s_{2} & 0 & c_{2}
\end{pmatrix}
\begin{pmatrix}
U_3 & 0_2\\
0_2 & U_4
\end{pmatrix},
\end{multline*}
where $s_{j}=\sin(a_{j}/2)$ and $c_{j}=\cos(a_{j}/2)$.
In particular, the Lie group $\GK$ is given in this basis by block-diagonal unitary transformations, 
where $0_2$ is the  $2{\times}2$-dimensional
zero matrix and $U_{1},U_{2}$ (and $U_{3},U_{4}$) are $2{\times}2$-dimensional unitary 
matrices such that the product of their determinants is one. The  
considered $\GK \GA \GK$ decomposition is equivalent to the cosine-sine decomposition
\cite{vL:1985,GvL:1989,PW:1994}.

{\it Remark~2.} In Ref.~\cite{KBG:2001}, a different Cartan decomposition is considered.
In that case, the subalgebra $\gk$ is given by the elements $-iS_{\mu}$ and $-iI_{\nu}$
and corresponds to unitary transformations on single qubits
of a coupled two-qubit system. Synthesizing unitary transformations on single qubits
is assumed in Ref.~\cite{KBG:2001} to take significantly less time, as compared to unitary transformations 
which interact between different qubits.

Since elements of $\GK$ can be synthesized in negligible time, we obtain as the main
result of this paper that the minimum time to synthesize any element $G \in \SU(4)$ is
the minimum value of $(|t_1| + |t_2|)/\omega_r^I$ such that $(t_1, t_2)^{T}$ is a pair satisfying Eq.~\eqref{eq:decompose}. We defer the proof of this fact to
Sec.~\ref{derivation}. Let us describe how to use the $\GK \GA \GK$ decomposition of 
$\G$, to 
synthesize an arbitrary transformation using only the 
generators $-iS_{\mu}$, $-i H^{\alpha}(\phi_{I})$, and $-iH_{0}$. 

The Lie algebra $\gk$ decomposes to $\gk_1 \oplus \gp_1$, where $\gk_1$ is a subalgebra, 
composed of operators $-iS_{\mu}$ and $-i2S_{\nu}I_{z}$, and $\gp_1$ is generated by
$-iI_{z}$ which commutes 
with all elements of $\gk_1$.
The Lie algebra $\gk_1$ can be further subdivided by a Cartan decomposition
$\gk_{1} = \gk_{2} \oplus \gp_2$. The subalgebra $\gk_{2}$ is generated by
the operators $-iS_{\mu}$,  and the subspace $\gp_{2}$ consists
of the operators $-i2S_{\mu}I_{z}$. Therefore, similar as in
Lemma~3, we obtain a decomposition of $\GK$: 

{\it Lemma~4.} Each element $K_j \in \GK$ can be decomposed as
$K_j = \exp(-i\tau_{2j-1}I_{z})L_{2j-1} \exp(-i\tau_{2j} 2S_{z}I_{z}) L_{2j}=$
\begin{gather} \label{eq:lab1}
\exp[-i(\tau_{2j-1}-\tau_{2j})I_{z}] L_{2j-1} \exp(-i \tau_{2j} H_0/J) L_{2j},
\end{gather}
where $\tau_{j}\in\R$ and $L_{j} \in \GK_{2}= \exp(\gk_2)$. 

Using an Euler angle decomposition (see, e.g., pp.~454--455
of Ref.~\cite{Shu:1993a}),
the elements $L_j\in\GK_2$ are given as
\begin{align}
\label{eq:lab2}
\nonumber L_j =& \exp(-i \theta_{j,1} S_z) \exp(-i \theta_{j,2} S_x) \exp(-i \theta_{j,3} S_z)\\
 =& \exp[-i (\theta_{j,1}+\theta_{j,3}) S_z] \exp[-i \theta_{j,2} R(\theta_{j,3})],
\end{align}
where $R(\theta_{j,3}) = S_x \cos \theta_{j,3} - S_y \sin \theta_{j,3}$.

\begin{figure}
\includegraphics{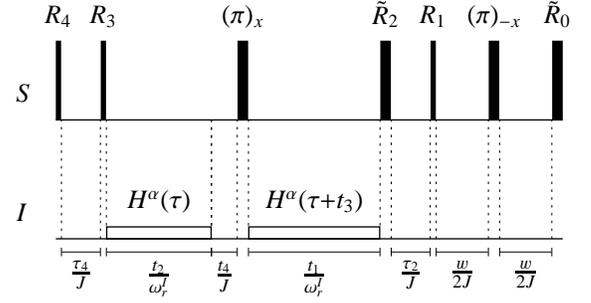}
\caption{\label{fig:pulse} The figure shows a canonical pulse sequence for synthesizing 
unitary transformations in the coupled qubit system.
Let $\tilde{R}_{2}=R_2  \exp(i\pi S_{x})$ and $\tilde{R}_0=\exp(-i v_0 S_{z})\exp(- i \pi S_x)$.
Since $1/J \ll 1/w_r^{I}$,
the length of the time intervals  $t_j/w_r^I$ is larger as depicted.  
 Refer to the text for details.}
\end{figure}

Similarly, any element $A$ of the subgroup $\GA$ can be written as
$ A = \exp[t_{1}(-iS^{\beta}I_{x})+t_{2}(-iS^{\alpha}I_{x})] = $
\begin{gather}
\nonumber \exp\left[-i\frac{t_{1}}{w_r^I} H^\beta(0)\right] e^{i t_3 I_z} e^{-i t_4 H_0/J} 
\exp\left[-i\frac{t_{2}}{w_r^I} H^\alpha(0)\right]= \\
e^{i t_3 I_z} e^{-i t_{1} H^\beta(t_3)/w_r^I } e^{-i t_4 H_0/J} 
e^{-i t_{2} H^\alpha(0)/w_r^I },\label{eq:lab3}
\end{gather}
for  $t_{3}=2Jt_{1}/w_r^I \mod 4\pi$ and $t_{4}=J(t_{1}-t_2)/w_r^I \mod 2\pi \geq 0$. 
This follows 
by substituting for expressions of $H_0$, $H^\alpha(\phi_{I})$, and $H^\beta(\phi_{I})$ (see Eqs.~\eqref{eq:H0}-\eqref{eq:beta}).
Combining Eqs.~\eqref{eq:lab1}-\eqref{eq:lab3}, 
a complete decomposition of an element $G \in \SU(4)$, can be written as
$K_1 A K_2=$
\begin{gather*}\label{eq:con}
e^{-i v_0 S_{z}} e^{-i w I_z} R_1
e^{-i \tau_{2} H_0/J} 
 R_2  \exp\left[-i \frac{t_{1}}{w_r^I} H^\beta(t_3+\tau)\right]\\ \times  e^{-i t_{4} H_0/J} 
 \exp\left[-i\frac{t_{2}}{w_r^I} H^\alpha(\tau)\right] R_3  e^{-i \tau_4 H_0/J}  R_4,
\end{gather*}
where all the  transformations $R_j$ operate on the fast qubit. 
In particular, we have
$R_4=\exp[-i\theta_{4,2} R(\theta_{4,3})]$, $R_3=\exp[-i \theta_{3,2} R(v_3)]$, $R_{2}=\exp[-i\theta_{2,2} R(v_2)]$,
$R_{1}=\exp[-i \theta_{1,2} R(v_1)]$,
$v_3=\theta_{3,3}+\theta_{4,1}+\theta_{4,3}$, $v_2=\theta_{2,3}+\theta_{3,1}+v_3$,
$v_1=\theta_{1,3}+\theta_{2,1}+v_2$,
$v_0=\theta_{1,1}+v_1$, $\tau=\tau_4-\tau_3$, and
$w=\tau_1-\tau_2+\tau_3-\tau_4-t_3$.
The time to produce $G$ is 
essentially $(t_{1}+ t_{2})/w_r^I$. Note that 
$\exp(-i w I_z)=$ 
$$ e^{- i \pi S_x} \exp[-i w H_0  /(2J)] e^{i \pi S_x} \exp[-i w H_0 /(2J)].$$  
Transformations on the fast qubit such as $\exp(-i v_0 S_{z})$ are significantly faster. 
Figure~\ref{fig:pulse} shows 
the canonical pulse sequence realizing any unitary transformation as a sequence 
of rotations under $-iH_0$, $-iH^\beta(\phi_{I})$, and $-iS_\mu$.

\section{Examples\label{examples}}
We introduce the unitary 
transformations $\mathrm{CNOT}[1,2]$,
$\mathrm{CNOT}[2,1]$, and $\mathrm{SWAP}$
which are given as  follows
$$
\begin{pmatrix}
1 & 0 & 0 & 0 \\
0 & 1 & 0 & 0 \\
0 & 0 & 0 & 1 \\
0 & 0 & 1 & 0 
\end{pmatrix},
\begin{pmatrix}
1 & 0 & 0 & 0 \\
0 & 0 & 0 & 1 \\
0 & 0 & 1 & 0 \\
0 & 1 & 0 & 0 
\end{pmatrix}, \text{ and }
\begin{pmatrix}
1 & 0 & 0 & 0 \\
0 & 0 & 1 & 0 \\
0 & 1 & 0 & 0 \\
0 & 0 & 0 & 1 
\end{pmatrix}.
$$
Let $c\in\{1,3,-1,-3\}$.
The elements of $\SU(4)$ corresponding to the transformation $\mathrm{CNOT}[2,1]$  are given by
$\exp[c\pi (-i2S_{x}I_{z}+iS_{x}+iI_{z})/2]$, which is equal to $\exp(ic\pi/4)\mathrm{CNOT}[2,1]$.
For $\mathrm{CNOT}[1,2]$ and $\mathrm{SWAP}$ we obtain the elements 
$\exp[c\pi (-i2S_{z}I_{x}+iS_{z}+iI_{x})/2]$ and $\exp[c\pi (i2S_{x}I_{x}+i2S_{y}I_{y}+i2S_{z}I_{z})/2]$,
which are equal to
$\exp(ic\pi/4)\mathrm{CNOT}[1,2]$ and $\exp(ic\pi/4)\mathrm{SWAP}$, respectively.
These different instances of unitary transformations result from the irrelevance of the global 
phase in quantum mechanics and can be described mathematically by multiplying with elements 
of the (finite) center of $\G$. The center consists of those elements which
commute with all elements of $\G$. To find the time-optimal control algorithm,
we may have to consider multiplying with different elements of the center.

As $\exp(i\pi/4)\mathrm{CNOT}[2,1]$ is an element of $\GK$, it 
takes negligible time to synthesize $\mathrm{CNOT}[2,1]$.
In strong contrast, $\exp(i\pi/4)\mathrm{CNOT}[1,2]$ is not contained in $\GK$.
Using the $\GK \GA \GK$ decomposition, both $\exp(i\pi/4)\mathrm{CNOT}[1,2]$ and  
$\exp(i\pi/4)\mathrm{SWAP}$ correspond to the same generator of $\GA$, given by 
$\pi(-iS^{\beta}I_{x})+0(-iS^{\alpha}I_{x})$, and
the minimum time to synthesize each of them is equal to $t_{\text{min}}=\pi$. This 
is still the optimal time if we consider 
to multiply with different elements of the center.

\begin{figure}
\includegraphics{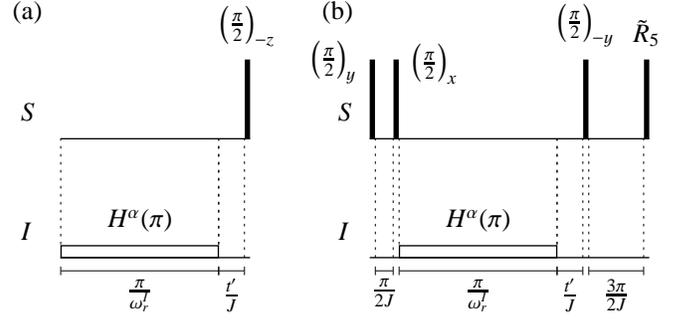}
\caption{\label{fig:pulse_exp} The figure shows the pulse sequences for synthesizing
the unitary transformations  (a) $\exp(i\pi/4)\mathrm{CNOT}[1,2]$
and (b) $\exp(i\pi/4)\mathrm{SWAP}$, where
 $\tilde{R}_{5}=  \exp(i\pi S_{z}/2) \exp(-i\pi S_{x}/2)$.
Since $1/J \ll 1/w_r^{I}$,
the length of the time intervals  $\pi/w_r^I$ is larger as depicted.
 Refer to the text for details.}
\end{figure}

We explicitly state the control algorithms:
The unitary transformation   $\exp(i\pi/4)\mathrm{CNOT}[1,2]$
is given by
\begin{gather*}
\exp(i \pi S_{z}/2)\exp(i\pi I_{z})
\exp(-i\pi S^{\alpha}I_{x})\exp(-i\pi I_{z})\\
=
\exp(i \pi S_{z}/2)
\exp( -i t' H_0 /J) \exp\left[-i \pi H^{\alpha}(\pi)/w_r^I\right],
\end{gather*}
where $t'=-\pi J / w_r^I \mod 2\pi\geq 0$.
Similarly, the unitary transformation $\exp(i\pi/4)\mathrm{SWAP}$ is given by
\begin{gather*}
e^{i\pi/4}\mathrm{CNOT}[2,1] e^{i\pi/4}
\mathrm{CNOT}[1,2] e^{-i\pi/4}
\mathrm{CNOT}[2,1] \\
= e^{i \pi S_{z}/2} e^{-i\pi S_x /2} e^{-i 3\pi H_0 /(2J)}  e^{i\pi S_y /2} 
e^{-i t' H_0 /J} \\ \times \exp\left[-i \pi H^{\alpha}(\pi)/w_r^I\right]
e^{-i\pi S_x /2}  e^{-i \pi H_0 /(2J)} e^{-i\pi S_y /2}.
\end{gather*}
The corresponding pulse sequences are given in Fig.~\ref{fig:pulse_exp}.

\section{Proof of time-optimality\label{derivation}}
In this section, we prove the time-optimality of
the given control algorithms in order
to synthesize unitary transformations in coupled fast and slow qubit system.
As expected, the maximal amplitude 
$\omega_{r}^I$ (see Eq.~\eqref{timescales}) determines the optimal time.

\subsection{The simple case}\label{simplecase}
All control algorithms, synthesizing a unitary transformation in time $t=\sum_{j} t_{j}$, can be written in the form
\begin{equation}\label{prod_new}
K_{n+1}' \exp[-it'_{n} H^\beta(\psi_n)] K_{n}' \cdots  K_{2}' \exp[-it'_{1} H^\beta(\psi_1)] K_{1}',
\end{equation}
where  $K_{j}' \in \GK$ take negligible time to be synthesized as compared to the evolution under
$H^{\beta}$, $t_j,\psi_{j}\in \R$, and $t'_j=t_j/w_r^{I}$. We can rewrite Eq.~\eqref{prod_new} as
\begin{equation}\label{prod}
K_{n+1} \exp[-it_{n} S^{\beta}I_{x}] K_{n} \cdots  K_{2} \exp[-it_{1}
S^{\beta}I_{x}] K_{1},
\end{equation}
where $K_{j}\in \GK$.
Equation~\eqref{prod} can be  rewritten as 
\begin{equation}\label{tildep}
\tilde{K}_{n+1} \exp(\tilde{p}_{n})  \cdots   \exp(\tilde{p}_{1}),
\end{equation}
where $\tilde{p}_{j}=\tilde{K}_{j}(-it_{j}S^{\beta}I_{x})\tilde{K}_{j}^{-1}$
and $\tilde{K}_{j}$ are suitable elements of $\GK$.
Observe that the elements $\tilde{p}_{j}$
are contained in $\gp$. This follows from the Campbell-Baker-Hausdorff formula (see, e.g.,
Appendix~B.4 of Ref.~\cite{Kna:2002}) and the fact that $[\gk, \gp] \in \gp $ (see Eq.~\ref{eq:Cartan}). 
It was shown in Ref.~\cite{KBG:2001}
that for all time-optimal control algorithms 
the elements $\tilde{K}_{j}$ can be chosen such that all $\tilde{p}_{j}$ commute.
Therefore, all $\tilde{p}_{j}$ belong to a maximal Abelian subalgebra inside $\gp$, and 
we can find one $K_0\in\GK$ such that
$K_0\tilde{p}_{j}K_0^{-1} \in \ga$ for all $j$. 
Using this result and results of Eq.~\eqref{weylorbit} below, we can rewrite
Eq.~\eqref{tildep} in the form
\begin{equation}\label{pEqu}
\bar{K}_{2} \exp(t_{n}p_{n})  \cdots   \exp(t_{1}p_{1})
\bar{K}_{1},
\end{equation}
where $p_{j}=\beta_{j}(-iS^{\beta}I_{x})+\alpha_{j}(-iS^{\alpha}I_{x})$,
$$(\beta_{j},\alpha_{j})^{T}\in\{(-1,0)^{T},(1,0)^{T},(0,-1)^{T},(0,1)^{T}\},$$ and $\bar{K}_{1},\bar{K}_{2}\in\GK$.
Equation~\eqref{pEqu}
can be simplified to 
\begin{equation}\label{optalgo}
\bar{K}_{2} \exp[\bar{\beta}(-iS^{\beta}I_{x})+ \bar{\alpha}(-iS^{\alpha}I_{x})] \bar{K}_{1},
\end{equation}
where $\bar{\alpha}=\sum_{j}\alpha_{j}t_{j}$ and
 $\bar{\beta}=\sum_{j}\beta_{j}t_{j}$.
Assume that the unitary transformation to be synthesized is given by one of 
its $\GK \GA \GK$ decompositions
$\bar{K}_{4} \exp[a_{1}(-iS^{\beta}I_{x})+ a_{2}(-iS^{\alpha}I_{x})] \bar{K}_{3}$, where
$a_{j} \in \R$ and $\bar{K}_{3},\bar{K}_{4}\in\GK$. We remark that the $\GK \GA \GK$ decomposition
is not unique, and we prove in Sec.~\ref{appwhichelements} that different $\GK \GA \GK$ decompositions
$\bar{K}_{6} \exp[a'_{1}(-iS^{\beta}I_{x})+ a'_{2}(-iS^{\alpha}I_{x})] \bar{K}_{5}$
correspond to all values $a'_{j}=a_{j}+2\pi z_{j}$, where $z_{j}\in \Z$ and 
$\bar{K}_{5},\bar{K}_{6}\in\GK$. We can choose $\bar{a}_{1}$
and $\bar{a}_{2}$ as those values  of $a'_{1}$ and $a'_{2}$ such that
$\abs{\bar{a}_{1}}+\abs{\bar{a}_{2}}$ is minimal. If $\abs{\bar{a}_{1}}+\abs{\bar{a}_{2}} >t$,
we cannot synthesize the  unitary transformation in time $t$ since all time-optimal
control algorithms are equal to Eq.~\eqref{optalgo} and 
$\abs{\bar{\alpha}}+\abs{\bar{\beta}}=\abs{\sum_{j} \alpha_{j} t_{j}}+
\abs{\sum_{j} \beta_{j} t_{j}}\leq \sum_{j} (\abs{\alpha_{j}}+\abs{\beta_{j}}) t_{j}=\sum_{j} t_{j}=t$.
For $\abs{\bar{a}_{1}}+\abs{\bar{a}_{2}}\leq t$, we can use the control algorithm
$\exp(-i \bar{a}_{1} S^{\beta}I_{x}) \exp(-i \pi S_{x}) \exp(-i \bar{a}_{2} S^{\beta}I_{x}) 
\exp(i \pi S_{x})$ to synthesize the unitary transformation in time 
$\abs{\bar{a}_{1}}+\abs{\bar{a}_{2}}$. 

\subsection{The general case}\label{generalcase}
Until now,  we have assumed  that in Eq.~\eqref{controlI},  $\omega_c^I = \omega_I -J$, i.e., we irradiate 
on the transition $\alpha\alpha \leftrightarrow \alpha\beta$. More generally, we can 
irradiate on both transitions $\alpha\alpha \leftrightarrow \alpha\beta$ and
$\beta\alpha\leftrightarrow\beta\beta$. 
Hence, we substitute Eq.~\eqref{controlI} by 
\begin{align*}
H_{I}^{\text{lab}}(t')=\, & 2 \omega_{r}^{I}(t') \{ b_2 \cos[(\omega_{I} - J)t' + \phi_2(t')]\\ +\, & b_1 \cos[(\omega_{I} + J)t' + \phi_1(t')]\}I_x,
\end{align*}
where $|b_1| + |b_2| \leq 1 $ (this ensures that the peak amplitude is $2\omega_r^I$). 
Transforming into a double rotating frame by 
\begin{equation*}
U_{\text{lab}}(t')=\exp(-it' \omega_{S} S_{z}) 
\exp[-i t' (w_{I} +J2 I_{z} S_{z} ) ] U_{\text{rot}}(t'),
\end{equation*}
the evolution under the control Hamiltonian for time $t'$ (with constant $\omega_r^I, \phi_1,\phi_2\in \R$) generates a net rotation $K''_1 \exp[-i t' \omega_r^I  (b_1 S^\beta I_p + b_2 S^\alpha I_q)]$, where $I_p=I_x \cos(\phi_1)+I_y \sin(\phi_1)$,
$I_q=I_x \cos(\phi_2)+I_y \sin(\phi_2)$, and $K''_1 \in \GK$.
This can be
rewritten as $K'_1 \exp(-it b) K'_2$, where $b=b_{1}(-iS^{\beta}I_{x})+b_{2}(-iS^{\alpha}I_{x})\in\ga$,
$K'_1,K'_2\in \GK$, and $t=t' \omega_r^I$.
Therefore, any control algorithm generates in time $t$, a transformation (written as
in Eq.~\eqref{prod})
\begin{equation*}
K_{n+1} \exp[-i  t_{n} b ] K_{n} \cdots  K_{2} \exp[-it_{1} b  ] K_{1},
\end{equation*}
where $t_j$ is given in units of $1/\omega_r^I$ and $\sum_j t_j = t$. 
This generalizes the case of $b_{1}=1$ and $b_{2}=0$, treated in Sec.~\ref{simplecase}.

Similarly as in Sec.~\ref{simplecase}, we obtain time-optimal control 
algorithms as in Eq.~\eqref{tildep}, where 
$\tilde{p}_{j}=\tilde{K}_{j} b  \tilde{K}_{j}^{-1}$ and $\tilde{K}_{j}$ are 
suitable elements of $\GK$.
Therefore, Eq.~\eqref{tildep} can be transformed
to Eq.~\eqref{pEqu}, where the commuting elements 
$p_j = \beta_{j} (-iS^{\beta}I_{x})+\alpha_{j}(-iS^{\alpha}I_{x})$
are contained in the Weyl orbit
$\We(b)=\{K b K^{-1}\colon\, K \in \GK\} \cap \ga$, i.e., $(\beta_{j}, \alpha_{j})^{T}$ is an element of the set
\begin{multline}\label{weylorbit}
\left\{
(b_{1},b_{2})^{T},(b_{1},-b_{2})^{T},(-b_{1},b_{2})^{T},(-b_{1},-b_{2})^{T}, \right. \\
\left.
(b_{2},b_{1})^{T},(-b_{2},b_{1})^{T},(b_{2},-b_{1})^{T},
(-b_{2},-b_{1})^{T}
\right\}.
\end{multline}
The Weyl orbit is induced  by the map $(K,b)\mapsto K b K^{-1}$,
where $b\in \ga$ and the elements $K\in\GK$ are 
\begin{multline}\label{Weyl}
\Big\{
\idf,\exp(-i \pi S^{\alpha}I_{z}),\exp(-i \pi S^{\beta}I_{z}),
\exp(-i \pi I_{z}),\\
\exp(-i\pi S_{x}),
\exp(-i\pi S_{x})\exp(-i \pi S^{\alpha}I_{z}),\\
\exp(-i\pi S_{x})\exp(-i \pi S^{\beta}I_{z}),
\exp(-i 2 \pi  S_{x} I_{z})
\Big\}.
\end{multline}
As before, Equation~\eqref{pEqu} can be simplified to Eq.~\eqref{optalgo}, and we obtain $\abs{\bar{\alpha}}+\abs{\bar{\beta}}\leq t (\abs{b_{1}}+\abs{b_{2}})$. Furthermore, $\max\{\abs{\bar{\alpha}},\abs{\bar{\beta}}\} \leq t \max\{\abs{b_{1}},\abs{b_{2}}\}$ holds. When the pairs $(a_1, a_2)^{T}$ and $(b_1, b_2)^{T}$ satisfy 
 $ \max\{ |a_1|, |a_2| \} \leq \max\{ |b_1|, |b_2| \}$ and $|a_1| + |a_2| \leq |b_1| + |b_2|$, then 
we say $(a_1, a_2)^T$ is $r$-majorized $(b_{1},b_{2})^{T}$, i.e., $(a_1, a_2)^T \prec_{r} (b_{1},b_{2})^{T}$. The notion of $r$-majorization is equivalent to
the condition that one element of $\ga$ is contained
in the convex closure of the Weyl orbit of another one
(for a proof see Appendix~\ref{appmaj}).

Given any unitary transformation $G\in\G$, let $t_{\text{opt}}$ be the smallest possible time such that 
\begin{equation}\label{equationT}
(a_1, a_2)^{T} \prec_{r} t_{\text{opt}} (b_1, b_2)^{T}
\end{equation}
and
$G = \bar{K}_{2} \exp[a_{1}(-iS^{\beta}I_{x})+a_{2}(-iS^{\alpha}I_{x})] \bar{K}_{1}$ with $\bar{K}_{j}\in \GK$. 
Again, the $\GK \GA \GK$ decomposition
is not unique, and different $\GK \GA \GK$ decompositions
correspond to all values $a'_{j}=a_{j}+2\pi z_{j}$, where $z_{j}\in \Z$
(see Sec.~\ref{appwhichelements}). Let us choose 
$a_{j}$ as an element of $[-\pi,\pi]$.
We prove in Appendix~\ref{apprefinedmaj} that for such a choice of $a_{j}$, 
the equation $(a_{1},a_{2})^{T}\prec_{r}(a_{1},a_{2})^{T}+2\pi(z_{1},z_{2})^{T}$
holds for all $z_{1},z_{2}\in\Z$. This implies that the smallest $t_{\text{opt}}$ in Eq.~\eqref{equationT}
can be achieved for $a_{1},a_{2}\in [-\pi,\pi]$. 

Then $G$ cannot be synthesized in time $t$ less than $t_{\text{opt}}$, as for such a control algorithm the 
equation $(\bar{\alpha},\bar{\beta})^{T} \prec_{r} t (b_{1},b_{2})^{T}$ would hold, and this would 
contradict the minimality of $t_{\text{opt}}$. In addition, $G$ can be synthesized in time $t$ greater 
than or equal $t_{\text{opt}}$: It follows from $(a_1, a_2)^{T} \prec_{r} t_{\text{opt}} (b_1, b_2)^{T}$ that 
$(a_1, a_2)^{T}$ is contained
in the convex closure of the Weyl orbit of $t_{\text{opt}} (b_1, b_2)^{T}$ (see Appendix~\ref{appmaj}) and
we can synthesize $G$ by convex combinations of elements of the Weyl orbit of $t_{\text{opt}}(b_1, b_2)^{T}$.

{\it Remark~3.}
Note, since $(b_1, b_2) \prec_{r} (1, 0)$, it follows that the minimum time to produce any unitary transformation can 
be obtained when all rf-amplitude is used to irradiate only on one nuclear transition 
(say $\alpha \alpha \leftrightarrow \alpha \beta$ as in Fig.~\ref{fig:energy}) as described earlier. 
This justifies our initial choice of irradiating only
on one nuclear transition.

\section{Conclusion\label{conclusion}}
In this paper, we presented
time-optimal control algorithms to synthesize arbitrary unitary transformations
for coupled fast and slow qubit system.
These control algorithms are applicable to electron-nuclear spin systems in pulsed EPR experiments
at high fields.
Explicit examples were given for $\mathrm{CNOT}$ and $\mathrm{SWAP}$.
Recently, controllability results have appeared for coupled electron-nuclear spin systems
at low fields \cite{Kha:2007,HJR:2007}, where it is shown that it is possible to synthesize
any unitary transformation on the electron-spin system by only manipulating the
electron. New methods need to be developed to obtain time-optimal control algorithms in these
settings. 

\begin{acknowledgments}
This work was supported by ONR 38A-1077404, AFOSR FA9550-05-1-0443,  and  NSF 0133673.
\end{acknowledgments}

\appendix

\section{Proofs}\label{proofs}

\subsection{Convex closure of  Weyl orbits\label{appmaj}}
Assume that $a=a_{1}(-iS^{\beta}I_{x})+a_{2}(-iS^{\alpha}I_{x})$
and $b=b_{1}(-iS^{\beta}I_{x})+b_{2}(-iS^{\alpha}I_{x})$ are elements of $\ga$.
We prove that $(a_1,a_2)^{T}$ is contained in the convex closure of the Weyl orbit of $(b_1,b_2)^{T}$ iff
$(a_{1},a_{2})^{T} \prec_{r} (b_{1},b_{2})^{T}$. 

Suppose $(a_1,a_2)^{T}$ is contained in the convex closure of the Weyl orbit of $(b_1,b_2)^{T}$. Assume that $\abs{b_1} \geq \abs{b_2}$. 
Then, $(a_1, a_2)^{T} = \sum_j w_j (b_{j,1}, b_{j,2})^{T}$, where $(b_{j,1}, b_{j,2})^{T}$ belongs to the set in Eq.~\eqref{weylorbit} ($w_j \geq 0$ and $\sum_j w_j = 1$). 
It follows that $\abs{b_{j,1}} \leq \abs{b_1}$ and $\abs{b_{j,2}} \leq \abs{b_1}$. Therefore,
$\abs{a_1} \leq \abs{b_1}$ and $\abs{a_2} \leq \abs{b_1}$, implying  
$\max\{ \abs{a_1}, \abs{a_2} \} \leq \max\{ \abs{b_1}, \abs{b_2} \}$. Also note,
$\abs{a_1} + \abs{a_2} \leq \sum_j w_j (\abs{b_{j,1}} + \abs{b_{j,2}}) = |b_1| + |b_2|$.

Suppose that $(a_1, a_2)^{T} \prec_{r} (b_1, b_2)^{T}$. The conditions
 $\max\{ \abs{a_1}, \abs{a_2} \} \leq \max\{\abs{b_1}, \abs{b_2} \}$ and $\abs{a_1} + \abs{a_2} \leq \abs{b_1} + \abs{b_2} $ are equivalent to $(\abs{a_1}, \abs{a_2})^{T}$ being
 weakly submajorized by $(\abs{b_1}, \abs{b_2})^{T}$. 
Thus, we obtain from Prop.~4.C.2.\ of Ref.~\cite{MO:1979} that $(\abs{a_1}, \abs{a_2})^{T}=$
\begin{gather*}
e_1 
(\abs{b_1}, \abs{b_2})^{T}
+e_2 
(\abs{b_2}, \abs{b_1})^{T}
+e_3
(\abs{b_1}, 0)^{T}\\
+e_4 
(0, \abs{b_1})^{T}
+e_5 
(\abs{b_2}, 0)^{T}
+e_6 
(0, \abs{b_2})^{T}
=\\
f_1
(\abs{b_1}, \abs{b_2})^{T}
+f_2 
(\abs{b_2}, \abs{b_1})^{T}
+f_3
(\abs{b_1}, -\abs{b_2})^{T}\\
+f_4
(-\abs{b_2}, \abs{b_1})^{T}
+f_5 
(\abs{b_2}, -\abs{b_1})^{T}
+f_6 
(-\abs{b_1}, \abs{b_2})^{T},
\end{gather*}
where $e_j \geq 0$, $\sum_j e_j =1 $, $f_1=e_1+(e_3+e_6)/2$,  $f_2=e_2+(e_4+e_5)/2$,
and $f_k=e_k/2$ for $k\in \{3,4,5,6\}$. In particular, we have that $f_j\geq 0$ (for all $j$) and $\sum_j f_j =1$.
It follows that  $(a_1,a_2)^{T}= (\epsilon_1 \abs{a_1}, \epsilon_2 \abs{a_2})^{T}
=$
\begin{gather*}
f_1
(\epsilon_3  b_1, \epsilon_4 b_2)^{T}
+f_2 
(\epsilon_5 b_2, \epsilon_6 b_1)^{T}
+f_3
(\epsilon_7 b_1, \epsilon_8 b_2)^{T}
\\
+f_4
(\epsilon_9 b_2, \epsilon_{10} b_1)^{T}
+f_5 
(\epsilon_{11} b_2, \epsilon_{12} b_1)^{T}
+f_6 
(\epsilon_{13} b_1, \epsilon_{14} b_2)^{T},
\end{gather*}
for appropriate 
choices of $\epsilon_j \in \{1,-1\}$.
We conclude the proof
by consulting Eq.~\eqref{weylorbit}.
A Lie-theoretic proof can be obtained by
following Thm.~2 of Ref.~\cite{ZGB:2004}.

\subsection{$\GK \GA \GK$ decomposition for elements of $\GA$\label{appwhichelements}}
We prove that the elements $\exp(a')\in \GA$ equal to
$K_{1} \exp(a) K_{2}$  are given by the elements
$(a_{1}',a_{2}')^{T}=(a_{1},a_{2})^T+2\pi(z_{1},z_{2})^T$, where 
$K_{j} \in \GK$, 
$a'=a'_1 (-iS^{\beta} I_x) + a'_2 (-iS^{\alpha} I_x)$, 
$a=a_1 (-iS^{\beta} I_x) + a_2 (-iS^{\alpha} I_x)$, and $z_{j} \in\Z$. 

We can choose $a'$ as $a'=K (a+k) K^{-1}$, where $K$ is an element of Eq.~\eqref{Weyl} and $k \in \{q \in \ga|\, \exp(q) \in \GK\}$
(cp.\ Ref.~\cite{ZGB:2004}, Lemma~2, and Ref.~\cite{DHHKL:2006}, Prop.~4).
Using the ansatz
$\exp[a''_{1}(-iS^{\alpha}I_{x})+a''_{2}(-iS^{\beta}I_{x})]=\idf$, where $a''_{1},a''_{2}\in \R$,
we obtain that $a''_{1},a''_{2} \in \{4\pi z\colon\, z \in \Z\}$.
It is a consequence of  Thm.~8.5, Chap.~VII, of Ref.~\cite{Hel:2001} that 
$\{q \in \ga|\, \exp(q) \in \GK\}$  is equal to $\{q_{1}(-iS^{\alpha}I_{x})+q_{2}(-iS^{\alpha}I_{x}):\,
q_{1},q_{2} \in \{2\pi z\colon\, z \in \Z\}\}$. This completes the proof. We remark
that $\exp[2\pi z_1 (-i S^{\beta} I_x) + 2 \pi z_2 (-i S^{\alpha} I_x)]=
\exp[2\pi z_1 (-i S^{\beta} I_z) + 2 \pi z_2 (-i S^{\alpha} I_z)]$ for all $z_j \in \Z$,
where $2\pi z_1 (-i S^{\beta} I_z) + 2 \pi z_2 (-i S^{\alpha} I_z)\in \gk$.

\subsection{Proof of a majorization relation\label{apprefinedmaj}}
We prove that 
$(a_{1},a_{2})^{T}\prec_{r}(a_{1},a_{2})^{T}+2\pi(z_{1},z_{2})^{T}$
holds for all $z_{1},z_{2}\in\Z$, if we assume that $a_{1},a_{2}\in [-\pi,\pi]$.

As the case $z_{1}=z_{2}=0$ is trivial, we assume
that $\abs{z_{1}}>0$ or $\abs{z_{2}}>0$. We obtain that
$\max\{\abs{a_{1}+2\pi z_{1}},\abs{a_{2}+2\pi z_{2}}\}\geq 2\pi - \pi = \pi \geq
\max\{\abs{a_{1}},\abs{a_{2}}\}$, and the first condition in the definition of
$r$-majorization is satisfied. The second condition
$\abs{a_{1}+2\pi z_{1}}+\abs{a_{2}+2\pi z_{2}}\geq\abs{a_{1}}+\abs{a_{2}}$ follows from the
fact that $\abs{a_{j}+2\pi z_{j}}\geq \abs{a_{j}}$ is always true. In particular,
this is trivial for $z_{j}=0$ and it is a consequence
of $\abs{a_{j}+2\pi z_{j}}\geq \abs{(\abs{2\pi z_{j}}-\abs{a_{j}})}\geq \pi \geq \abs{a_{j}}$
in all other cases. The result follows.



\end{document}